\documentstyle[prl,aps,floats,eclepsf]{revtex}
\begin{document}
\draft 
\wideabs{
\title{Chemical potential shift in lightly-doped to overdoped \\
Bi$_2$Sr$_2$Ca$_{1-x}${\it R}$_{x}$Cu$_2$O$_{8+y}$ ({\it R} = Pr, Er)}
\author{N.~Harima and A.~Fujimori}
\address{Department of Physics and Department of Complexity Science
  and Engineering, University of Tokyo, Bunkyo-ku, Tokyo 113-0033, Japan}
\author{T.~Sugaya and I.~Terasaki}
\address{Department of Applied Physics, Waseda University, Tokyo
169-8555, Japan}
\date{\today} 
\maketitle

\begin{abstract}
 We have studied the chemical potential shift in the high-temperature superconductor Bi$_2$Sr$_2$Ca$_{1-x}${\it R}$_{x}$Cu$_2$O$_{8+y}$ ({\it R} = Pr, Er), where the hole concentration is varied from 0.025 to 0.17 per Cu, by precise measurements of core-level photoemission spectra. The result shows that the shift becomes slow in the underdoped region as in the case of La$_{2-x}$Sr$_{x}$CuO$_{4}$ (LSCO) but the effect is much weaker than in LSCO. The observed shift in the present system can be relatively well explained by numerical results on the doped two-dimensional Hubbard model, and suggests that the change of the electronic structure induced by hole doping is less influenced by stripe fluctuations than in LSCO.
\end{abstract}
}

\narrowtext

\section{introduction}

Since the discovery of the high-temperature superconductors, one of the most important but unanswered questions has been how the electronic structure evolves from the antiferromagnetic insulator (AFI) to the superconductor (SC) as a function of doped hole concentration. In the case of La$_{2-x}$Sr$_x$CuO$_4$ (LSCO), recent angle-resolved photoemission (ARPES) studies \cite{LSCO_ARPES,LSCO_ARPESwhole} have shown that in underdoped samples, the chemical potential $\mu$ is pinned at $\sim$0.4 eV above the top of the lower Hubbard band (more precisely, top of the Zhang-Rice singlet band) and spectral weight is transferred from the lower Hubbard band to near $\mu$ with hole doping. Also, it has been found from the photoemission measurements of core levels that the chemical potential does not move with hole doping in the underdoped region~\cite{LSCO_mu}. These observations indicate a breakdown of the rigid-band picture and are suggestive of a dramatic reorganization of the electronic structure upon hole doping. In fact, the suppression of the chemical potential shift has been attributed to the strong stripe fluctuations in LSCO~\cite{YamadaLSCO} because the charge stripes can be viewed as a kind of microscopic phase separation which will pin the chemical potential. More recently, it has been found that the chemical potential in the electron-doped superconductor Nd$_{2-x}$Ce$_x$CuO$_4$ (NCCO) shows a monotonous shift without any sign of suppression, indicating a more rigid-band-like behavior~\cite{Harima}. The absence of a suppression of the shift in NCCO is consistent with the absence of stripe fluctuations in this system~\cite{yamada}.

As for Bi$_2$Sr$_2$CaCu$_2$O$_{8+y}$ (BSCCO), which has been most extensively investigated by ARPES~\cite{d_wave}, core-level shifts were studied on Bi$_2$Sr$_2$Ca$_{1-x}$Y$_{x}$Cu$_2$O$_{8+y}$, where hole concentration was varied with Y substitution~\cite{BSCCO_mu}. The deduced chemical potential showed an abrupt, large downward shift of ${\sim}0.7$ eV upon hole doping in the lightly doped region, which implies that the chemical potential moves from the in-gap region of the parent the insulator to the top of the lower Hubbard band upon hole doping. However, a subsequent study by Tjernberg {\it et al.}~\cite{BSCCO_core1} showed a different doping dependence of the core-level shifts. Therefore, more systematic studies over a wide concentration range are required to elucidate the intrinsic behavior of the chemical potential shift. So far, it has been difficult to prepare heavily underdoped BSCCO samples with good quality. Recently, high quality crystals of heavily underdoped BSCCO were systematically synthesized by rare-earth ($R$) substitution for Ca and the doping dependence of thermodynamic and transport properties have been studied \cite{Terasaki,sample}. 

In this paper, we report on a core-level photoemission study of the chemical potential shift ${\Delta}{\mu}$ in those Bi$_2$Sr$_2$Ca$_{1-x}${\it R}$_{x}$Cu$_2$O$_{8+y}$ crystals as a function of doped hole concentration. The chemical potential shift was estimated from the core-level shifts as in the previous studies \cite{LSCO_mu,Satake,Harima} utilizing the fact that the energies of core levels are measured reference to $\mu$ in photoemission experiments. 

\section{experiment}

Single crystals of Bi$_2$Sr$_2$Ca$_{1-x}R_{x}$Cu$_2$O$_{8+y}$ ($R$ = Pr, Er) were grown by the self-flux method and the x-ray diffraction pattern showed no trace of impurity phases. Details of sample preparation are given in \cite{sample}. The hole concentration ${\delta}$ per Cu atom was estimated from the empirical relationship between ${\delta}$ and the room-temperature thermopower \cite{thermopower}. Errors in $\delta$ thus estimated are less than $\pm$10 \% of $\delta$. The hole concentration and the critical temperature were ${\delta} = $0.17, 0.135 and 0.1 and $T_c=$ 86, 88 and 48 K for the $x_{\rm Pr}=$0.1, 0.25 and 0.43 samples, respectively, and ${\delta}=$0.135 and $T_c=$87 K for the $x_{\rm Er}=$0.1 sample. The $x_{\rm Er}=$0.5 and 1.0 samples were antiferromagnetic insulators and the hole concentration was ${\delta}=$0.05 and 0.025, respectively. The determination of the N\'{e}el temperatures $T_N$ of the present samples were difficult because of the Er$^{3+}$ local spins, but we may estimate $T_N{\sim}$50 K and 230 K for $x_{\rm Er}=$ 0.5 and 1.0 from comparison with $T_N$ values of Bi$_2$Sr$_2$Ca$_{1-x}$Y$_{x}$Cu$_2$O$_{8+y}$~\cite{Nishida}.

X-ray photoemission spectroscopy (XPS) measurements were carried out using a Mg $K{\alpha}$ source (${\it h}{\nu}=1253.6$~eV) and a VSW hemispherical analyzer. The energy resolution was about $0.8$~eV, which was largely due to the width of the photon source. The samples were cleaved {\it in situ} to obtain clean surfaces and measured at $\sim$ 80 K. The base pressure in the analyzer chamber was $\sim 1\times 10^{-10}$~Torr. All the spectra presented here were taken within four hours after cleaving and no change was observed in the spectra during the measurement. In order to avoid sample degradation or contamination, the samples were cooled to about $80$~K. In XPS measurements, high voltages of ${\sim}1$ kV have to be applied to the electron energy analyzer to decelerate photoelectrons, and it is usually difficult to measure the binding energies with an accuracy of $< 100$ meV. In order to overcome this difficulty, we monitored the applied voltages directly, and confirmed that the uncertainty could be reduced to less than $10$ meV~\cite{Harima}. Furthermore, in order to eliminate other unexpected causes of errors, we measured the $x_{\rm Pr}=$ 0.25 and $x_{\rm Er}=$ 0.1 sample as a reference following the measurement of each sample. However, the line shapes were not always identical between the different samples, which sometimes results in uncertainties up to $\pm$100 meV in determining the shifts (as indicated by error bars in Figs. 2, 4 and 5). 

\section{results and discussion}

Figure~1 shows the Sr $3d$, O $1s$ and Bi $4f$ core-level XPS spectra of Bi$_2$Sr$_2$Ca$_{1-x}${\it R}$_x$Cu$_2$O$_{8+y}$ ({\it R} = Pr, Er). For the Pr samples, the line shape of the Sr $3d$ level was almost identical between the different compositions, whereas for the Er samples, it varied slightly with composition on the high binding energy side of the peak. In order to estimate the energy shift, we used the position of the slope on the lower binding energy side of the peak for the Pr samples with higher precision and the peak positions for the Er samples (although only the peak positions are indicated in the figure for clarity). In Fig.~2, the Sr $3d$ peak position is plotted as a function of hole concenration ${\delta}$ per Cu atom.  The figure shows that the peak moves upwards monotonously with increasing ${\delta}$. 

The line shape of the Bi 4{\it f} level shown in Fig.~1(b) is mostly identical between the different compositions, and we could use the peak positions to evaluate the energy shift rather accurately. In the region ${\delta}\ge 0.1$, the peak is shifted upwards as ${\delta}$ increases, parallel to the Sr 3{\it d} core level. However, in the region ${\delta}<0.1$, the Bi 4{\it f} level moves downwards as ${\delta}$ increases, unlike Sr $3d$. One can understand this tendency if the mean valence of Bi decreases in the ${\delta}<0.1$ region and the Bi 4{\it f} level shows a so-called ``chemical shift''. Such a chemical shift cannot occur in the Sr $3d$ core level since the valence of Sr$^{2+}$ should be stable in metal oxides.

As for the O 1{\it s} spectra [Fig.~1(c)], the line shape is almost identical between the ${\delta}=0.025$ and ${\delta}=0.05$ samples, but becomes broader and somewhat asymmetric as ${\delta}$ increases to $\ge$0.1. This behavior is consistent with the previous study, where the O 1{\it s} core-level spectra were decomposed into several components \cite{BSCCO_core1,BSCCO_core2}. Note that in going from ${\delta}= 0.05$ to 0.1, where the O $1s$ width increases, the Bi 4{\it f} level moves to the direction opposite to Sr $3d$. Here, it should be remembered that there are at least three kinds of oxygen sites in Bi$_2$Sr$_2$Ca$_{1-x}${\it R}$_x$Cu$_2$O$_{8+y}$, namely, oxygen in the Bi-O layer, that in the SrO layer and that the CuO$_{2}$ layer. Then, considering the peculiar shift of the Bi $4f$ level, one may speculate that the broadening is caused by different shifts of the O 1{\it s} core levels from the diferent oxygen sites in the small $\delta$ region. 

In order to separately deduce the shifts of the O 1{\it s} signals from the different layers, we fitted the O $1s$ peak to a superposition of three components as shown in Fig.~3(a). Here, components A, B and C are assigned to oxygens from the CuO$_{2}$, SrO and Bi-O layers, respectively, following the previous study \cite{BSCCO_core2}. In the fitting procedure, we assumed the peak intensity ratio to be A:B:C = 1.2:1.0:1.8 taking into account the amount of excess oxygens $y =$ 0.15-0.2 in the Bi-O layers~\cite{crystal} as well as the mean free paths of photoelectrons escaping from the outermost Bi-O layer. First, we fitted the broad O 1{\it s} peak of the $\delta=0.135$ sample under the constraint that the widths of the three components were identical, and obtained the energy of each component and the common line width~\cite{comment}. Then, we fitted the O 1{\it s} peaks for the other compositions using the same widths. The shifts of the three components thus deduced are plotted in Fig.~3(b). One notices that as ${\delta}$ increases, the CuO$_{2}$ and SrO components (A and B) move upwards like the Sr $3d$ peak, while the Bi-O component (C) moves to downwards like the Bi $4f$ peak, leading to the broadening of the O 1{\it s} peak.

As for the Cu 2{\it p} core-level spectra (not shown), the peak becomes broader as ${\delta}$ increases in the Er samples. In the Pr samples, the Pr 3{\it d} core level overlaps with the Cu 2{\it p} core levels and therefore, it was difficult to identify the doping dependence of the line shape. Instead, we used the peak position to crudely estimate the shift of the Cu 2{\it p} level. In Fig.~2, we have plotted the shift of the Cu $2p$ peak positions for the Pr and Er samples. Thus, we find that the Cu 2{\it p} core level moves downards with $\delta$. This behavior can be understood if the chemical shift, which is caused by the creation of a ``Cu$^{3+}$" component on the higher binding energy side of the Cu$^{2+}$ main component, overwhelms the effect of the chemical potential shift, as in the case of LSCO~\cite{LSCO_mu}. 

In order to deduce the chemical poential shift from the set of the core-level shift data, we note that the shift ${\Delta}E$ of a core-level energy measured relative to $\mu$ is given by ${\Delta}E = -\Delta\mu + K\Delta{Q} + \Delta{V_{M}} + {\Delta}E_{R}$, where $\Delta\mu$ is the change in the chemical potential, $K\Delta{Q}$ is the chemical shift, $\Delta{V_{M}}$ is a shift due to a change in the Madelung potential, and ${\Delta}E_{R}$ is the change in the core-hole screening~\cite{Hufner}. It has been demonstrated that if the shifts of the metal and oxygen core levels are the same, $K\Delta{Q}$ and $\Delta{V_{M}}$ are negligibly small~\cite{LSCO_mu,Satake,Harima}. Core-hole screening by conduction electrons can also be excluded from the main origin of the core-level shifts in transition-metal oxides~\cite{LSCO_mu,Harima}. As mentioned above, in the region ${\delta}>0.1$, the Bi 4{\it f}, Sr 3{\it d} and O 1{\it s} core levels show similar behaviors and therefore we can safely conclude that the shifts are primarily caused by the chemical potential shift. In the region ${\delta}<0.1$, the doping dependences of the core levels are complicated, most likely due to charge transfer from the CuO$_2$ layers (CuO$_2$-Ca$_{1-x}R_x$-CuO$_2$ blocks) to the Bi-O layers with decreasing $\delta$ and the resulting decrease of the Bi valence. Considering the crystal chemistry of BSCCO, the SrO layer is supposed to have fixed charges and the Sr 3{\it d} level would be least influenced by the charge transfer, and we assume that the shift of the Sr 3{\it d} level most faithfully reflects the chemical potential shift in the entire hole concentration region. Indeed, the shift of the O $1s$ peak from the SrO layer as well as that from the CuO$_2$ layer deduced from the line-shape analysis show nearly the same behavior and is most probably caused by the chemical potential shift. 

Figure~4 shows the chemical potential shift $\Delta\mu$ relative to $\delta = 0$ thus deduced for the Er- and Pr-substituted BSCCO samples. The shift is slow in the underdoped regime, and becomes faster as the hole concentration increases. The depression of the chemical potential shift in the underdoped region, i.e., $|{\partial}{\mu}/{\partial}\delta|$ ${\rightarrow}$ $0$ as ${\delta}$ ${\rightarrow}$ $0$, has been predicted by numerical studies of the two-dimensional Hubbard model~\cite{Furukawa,Dagotto} and $t$-$J$ model~\cite{Prelovsek}. In the Monte Carlo simulations~\cite{Furukawa}, the calculated $\Delta\mu$ follows ${\Delta}{\mu}$ ${\propto}$ $-{\delta}^2$. The measured $\Delta\mu$ in BSCCO can be relatively well fitted to $-{\delta}^2$ as shown in Fig.~4. Note that this behavior has not been predicted by mean-field theories such as the Gutzwiller approximation and the dynamical mean-field approximation, which predict that $|{\partial}{\mu}/{\partial}\delta|$ ${\rightarrow}$ ${\infty}$ as ${\delta}$ ${\rightarrow}$ 0 \cite{Kotliar}.

In Fig.~5, we compare $\Delta\mu$ for BSCCO with that for LSCO. Both $\Delta\mu$ curves show similar doping dependences in the sense that the shift is slow in the underdoped region and fast in the overdoped region. However, ${\Delta}{\mu}$ is strongly suppressed in underdoped LSCO as if the chemical potential is pinned by some levels, while such a pinning-like behavior is not evident or much weaker in BSCCO (although a systematic deviation from the $-\delta^2$ behavior due to a weak $\mu$ pinning cannot be excluded from Fig.~4). Further, the shift in BSCCO is faster than that in LSCO in the entire hole concentration region. These differences indicate that the change in the electronic structure from the AFI to the SC with hole doping is different at least quantitatively between the two systems. In LSCO, it has been found by an ARPES study that $\mu$ is located in states which are created within the band gap of the parent insulator La$_2$CuO$_4$. ARPES data of underdoped BSCCO, on the other hand, appear to show a gradual upwards shift of the lower Hubbard band [around {\bf k}$=(\pi,0)$] with hole doping~\cite{Laughlin}, indicating no pinning behavior. It is highly desired to perform a systematic ARPES study lightly-doped to underdoped BSCCO. Because the pinning behavior of the chemical potential in LSCO has been attributed to charge fluctuations of the stripe-type~\cite{LSCO_mu}, the weakness or absence of the pinning behavior in BSCCO indicates that stripe fluctuations, if exist, are weaker in this system. 

\section{conclusion}

We have measured the chemical potential shift in BSCCO. The measured shift becomes slow in the underdoped region but is better fitted to $-{\delta}^2$ predicted by the numerical study of the two-dimensional Hubbard model than in the case of LSCO. The difference from LSCO is attributed to the weakness of stripe-type charge fluctuations in BSCCO. 

\section*{acknowledgment}

The authors would like to thank J. Matsuno for useful advice and A. Ino and T. Mizokawa for discussions. This work was supported by a Grant-in-Aid for Scientific Research in Priority Area ``Novel Quantum Phenomena in Transition Metal Oxides'' and a Special Coordination Fund for the Promotion of Science and Technology from the Ministry of Education, Culture, Sports, Science and Technology and New Energy and Industrial Technology Development Organization (NEDO).

\begin{figure}[htb]
\begin{center}
\epsfile{file=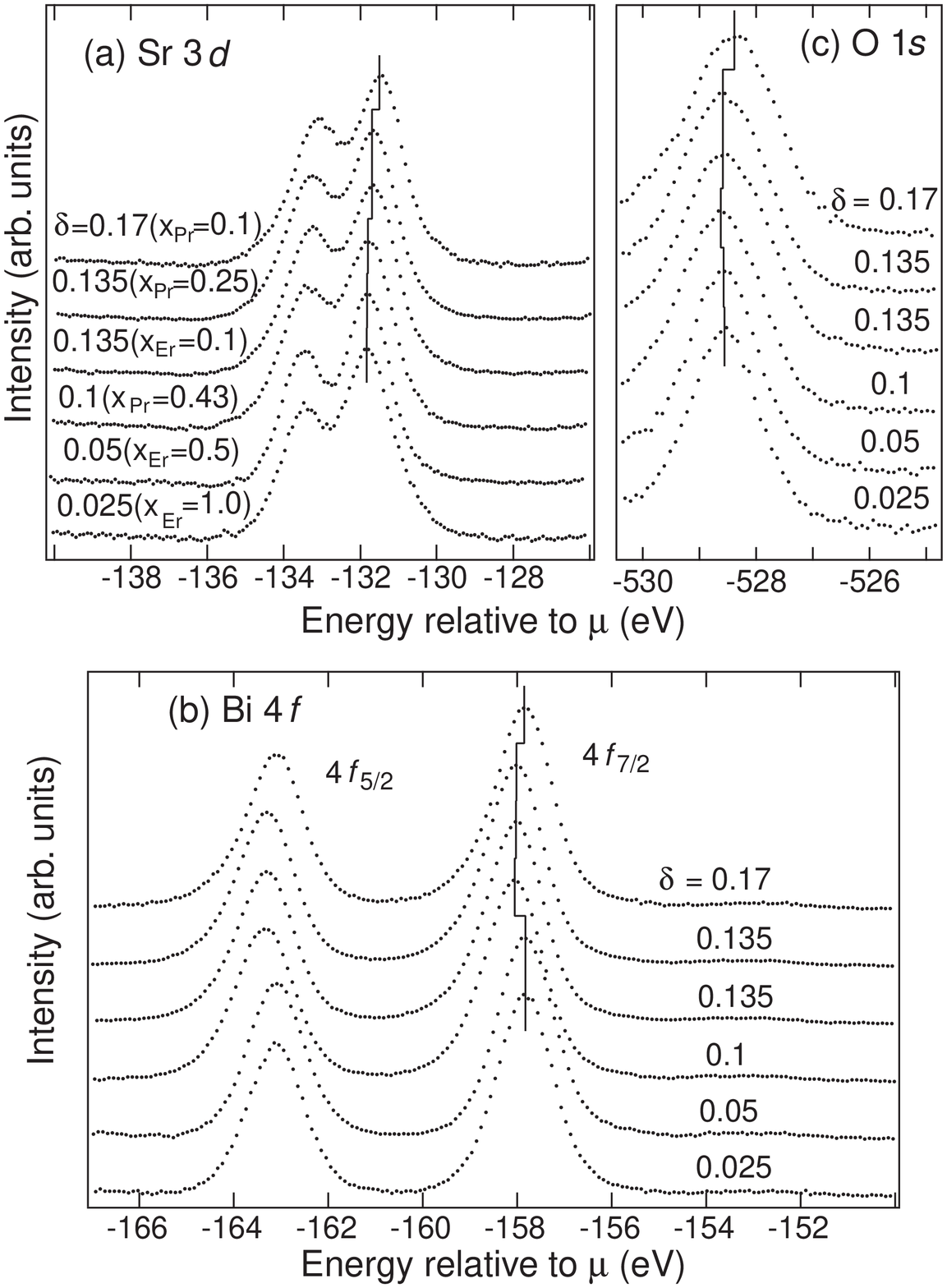,width=90mm}
\caption{Core-level XPS spectra of Er- and Pr-substituted BSCCO. (a): Sr 3{\it d}, (b): Bi 4{\it f}, (c): O 1{\it s}. In each spectrum, the vertical bar denotes the peak position, and $\delta$ denotes the hole concentration per Cu atom.}
\end{center}
\end{figure}

\begin{figure}[htb]
\begin{center}
\epsfile{file=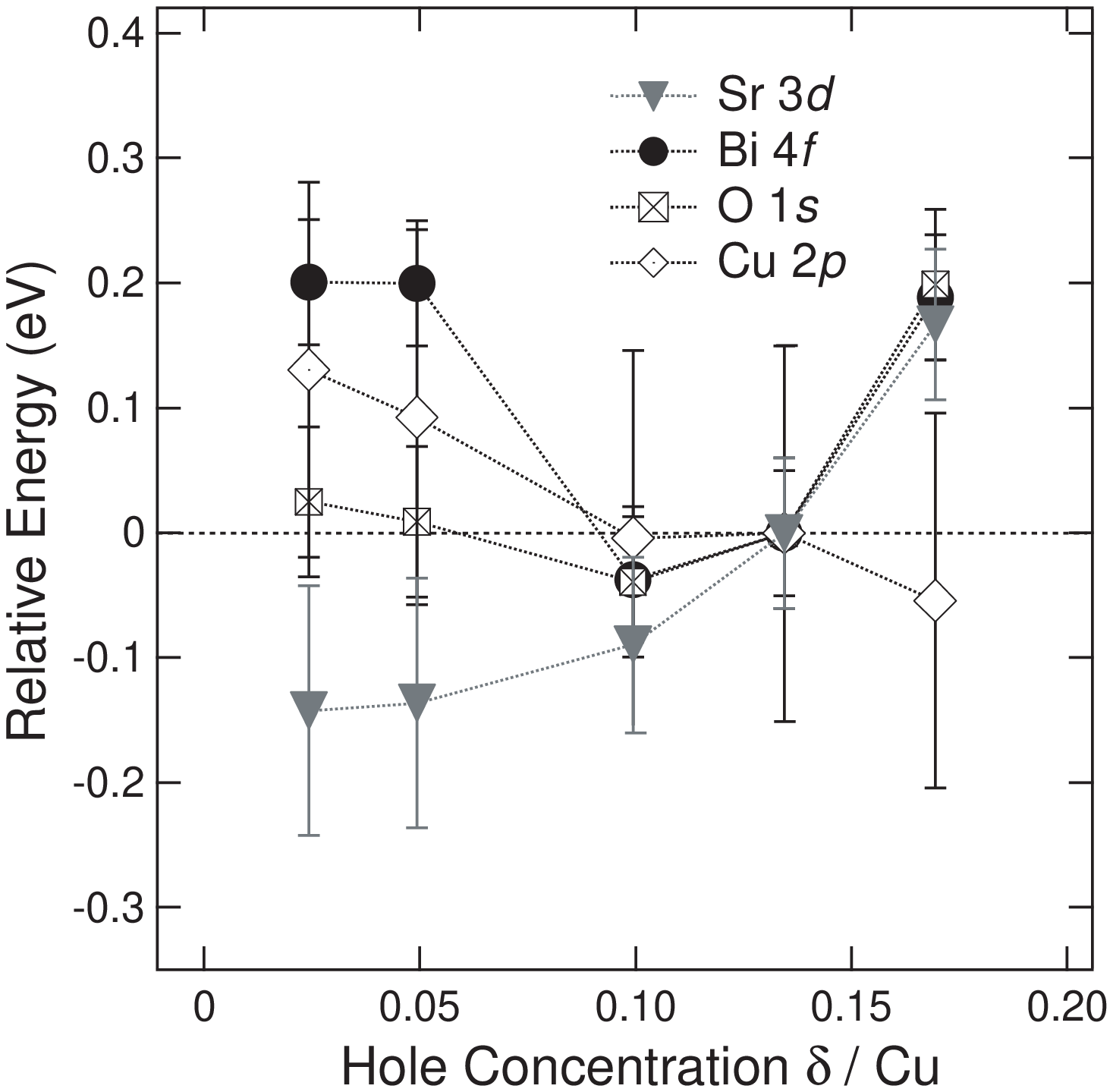,width=70mm}
\caption{Energy of each core level (relative to $\delta=$ 0.135) for Er- and Pr-substituted BSCCO as a function of hole concentration ${\delta}$.}
\end{center}
\end{figure}

\begin{figure}[htb]
\begin{center}
\epsfile{file=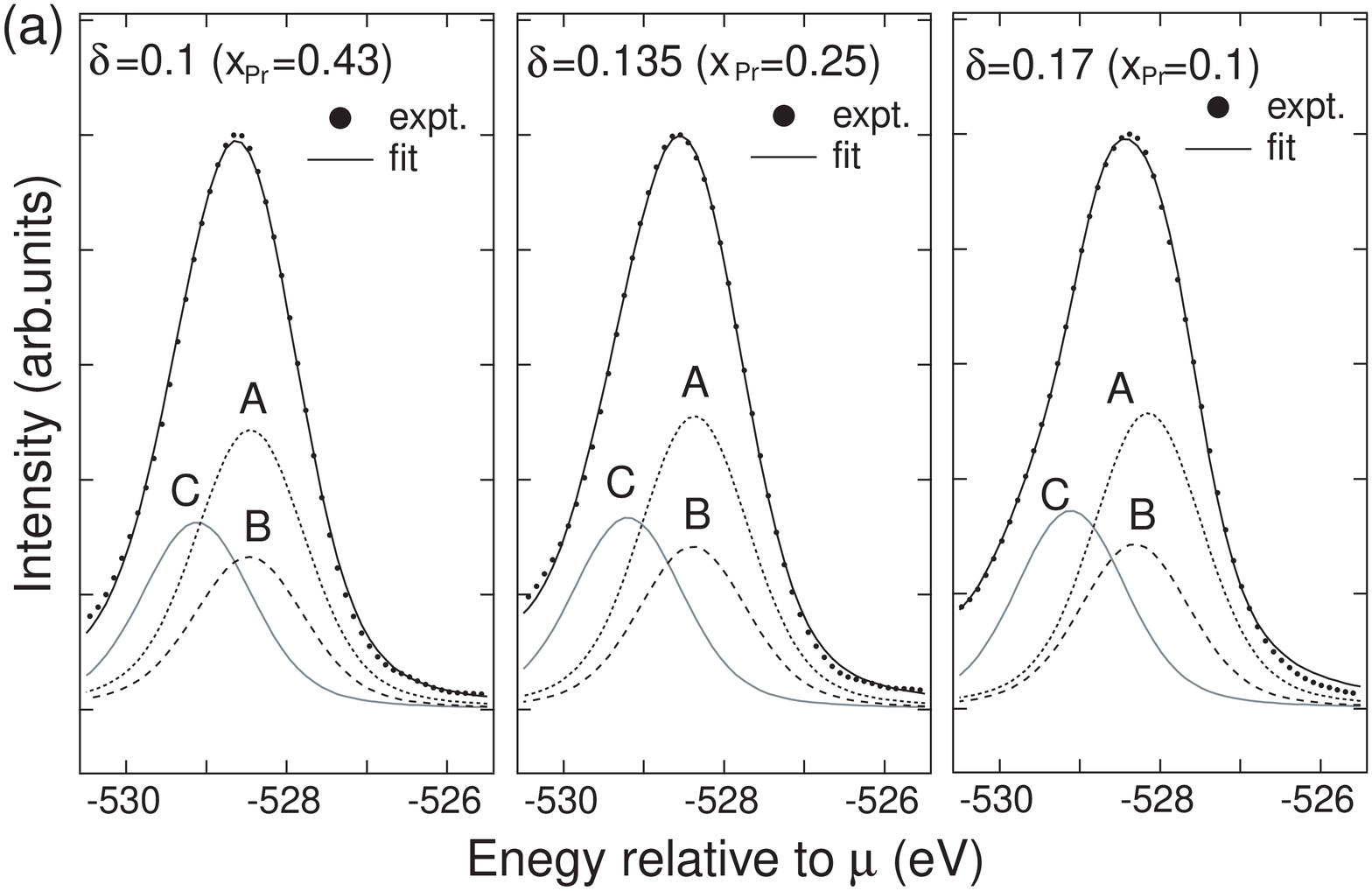,width=85mm}
\epsfile{file=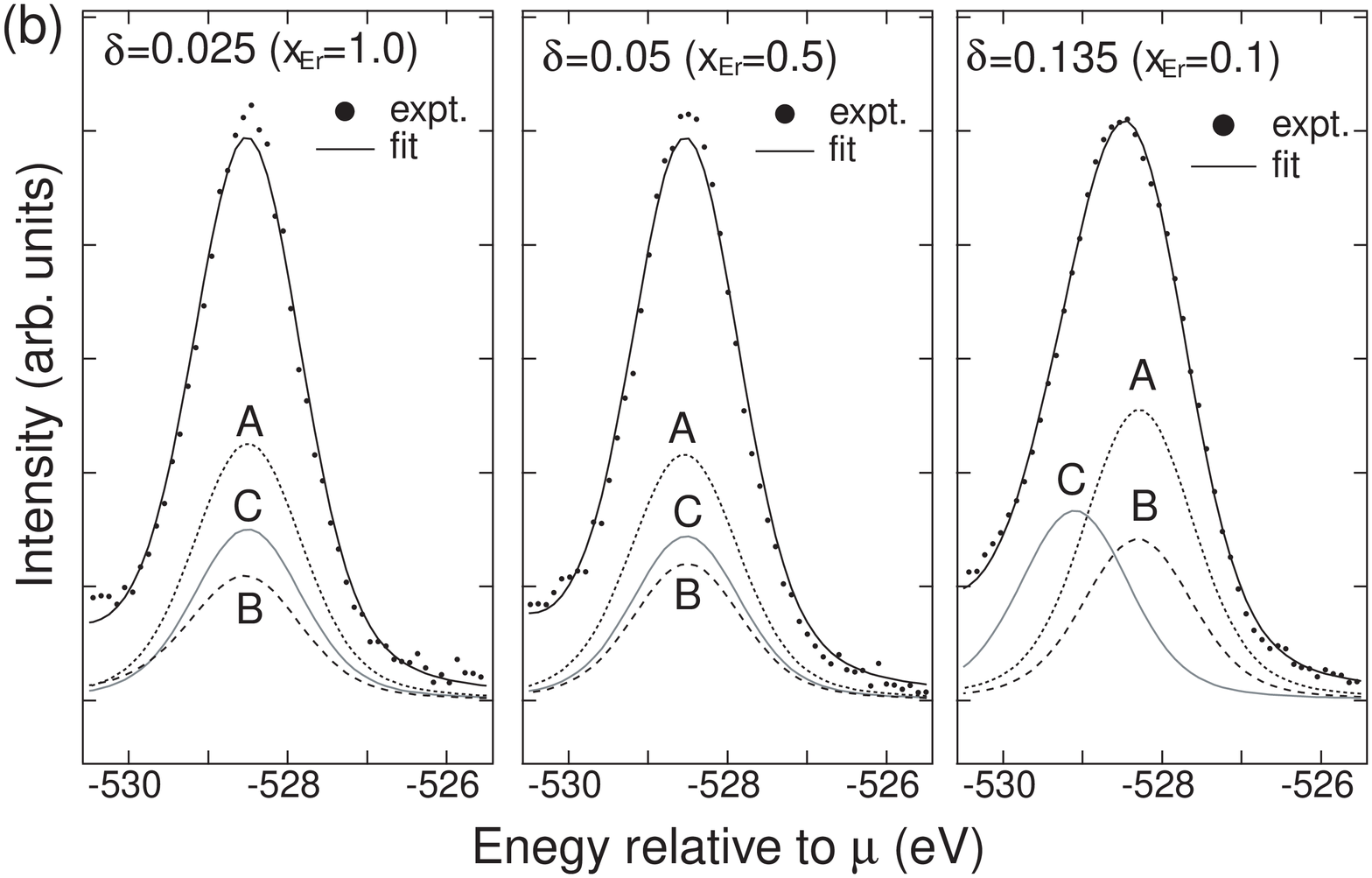,width=85mm}
\epsfile{file=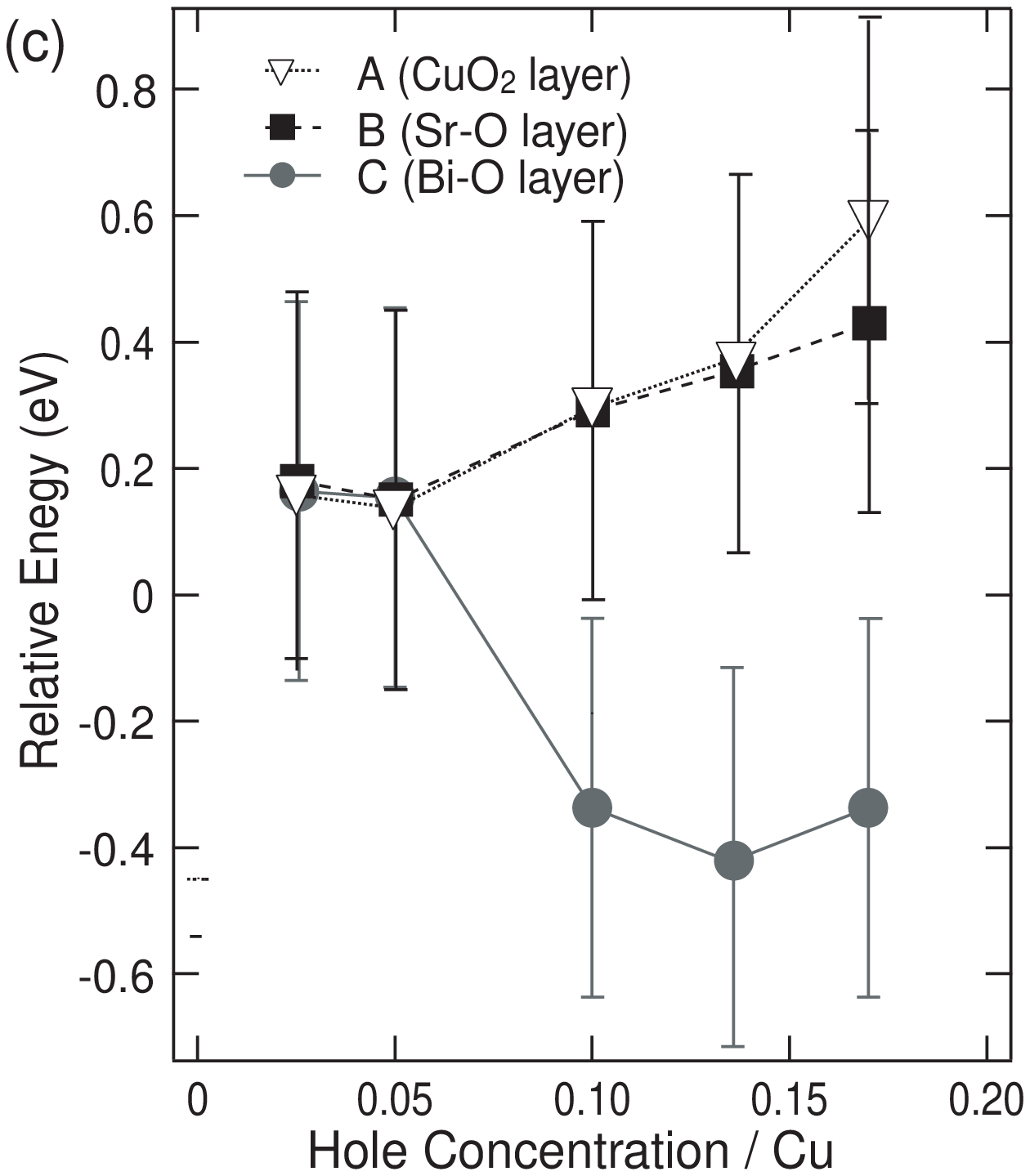,width=60mm}
\caption{O 1{\it s} core-level spectra of Pr- (a) and Er-substituted (b) BSCCO fitted to three components, which correspond to oxygens in the CuO$_2$ (A), SrO (B) and Bi-O (C) layers. (c) Energies of the three O $1s$ core-level components.}
\end{center}
\end{figure}

\begin{figure}[htb]
\begin{center}
\epsfile{file=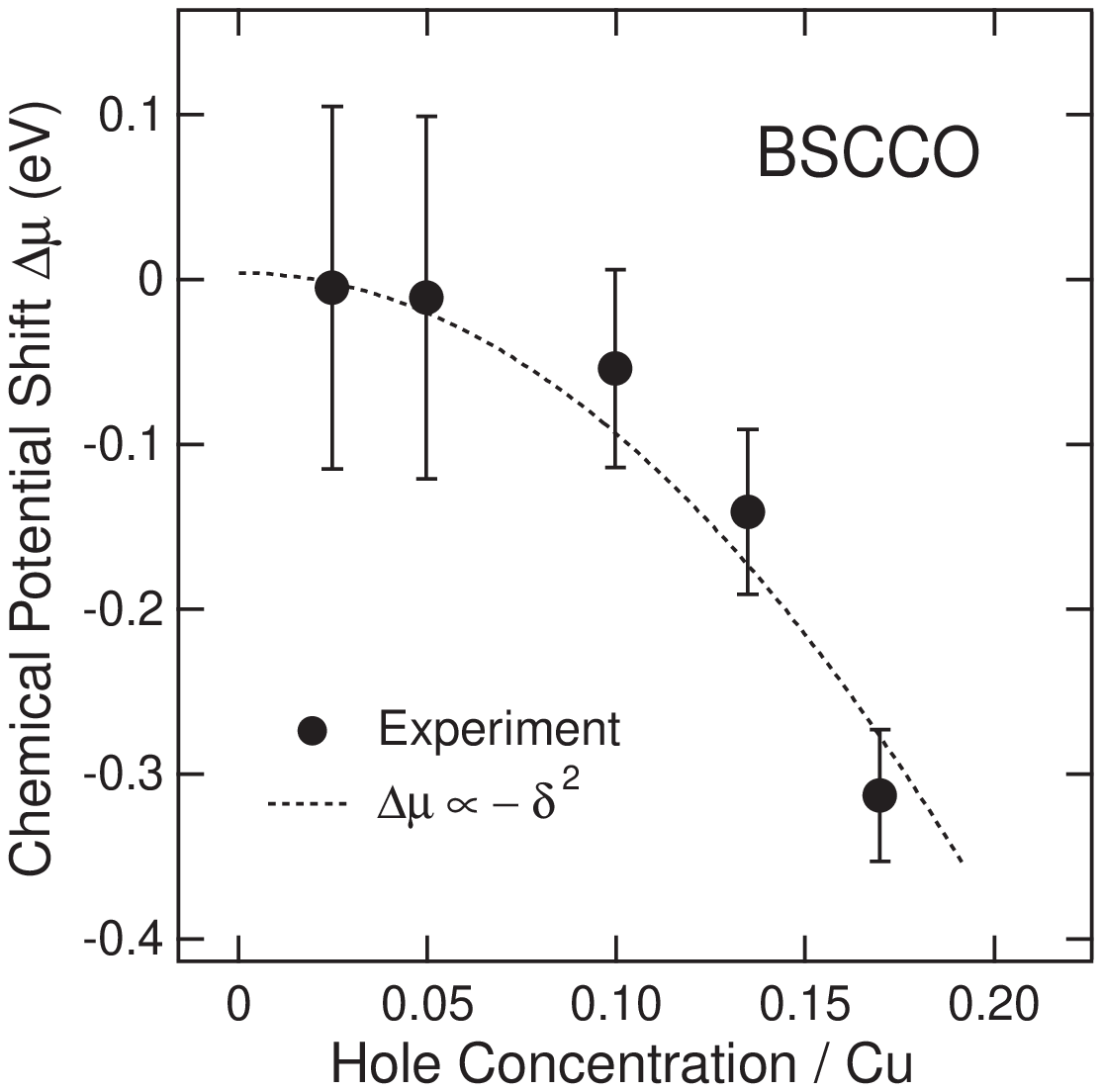,width=65mm}
\caption{Chemical potential shift ${\Delta}{\mu}$ in Er- and Pr-substituted BSCCO as a function of doped hole concentration ${\delta}$. A fit of ${\Delta}{\mu} \propto {\delta}^2$ is shown by a dashed curve.}
\end{center}
\end{figure}

\begin{figure}[htb]
\begin{center}
\epsfile{file=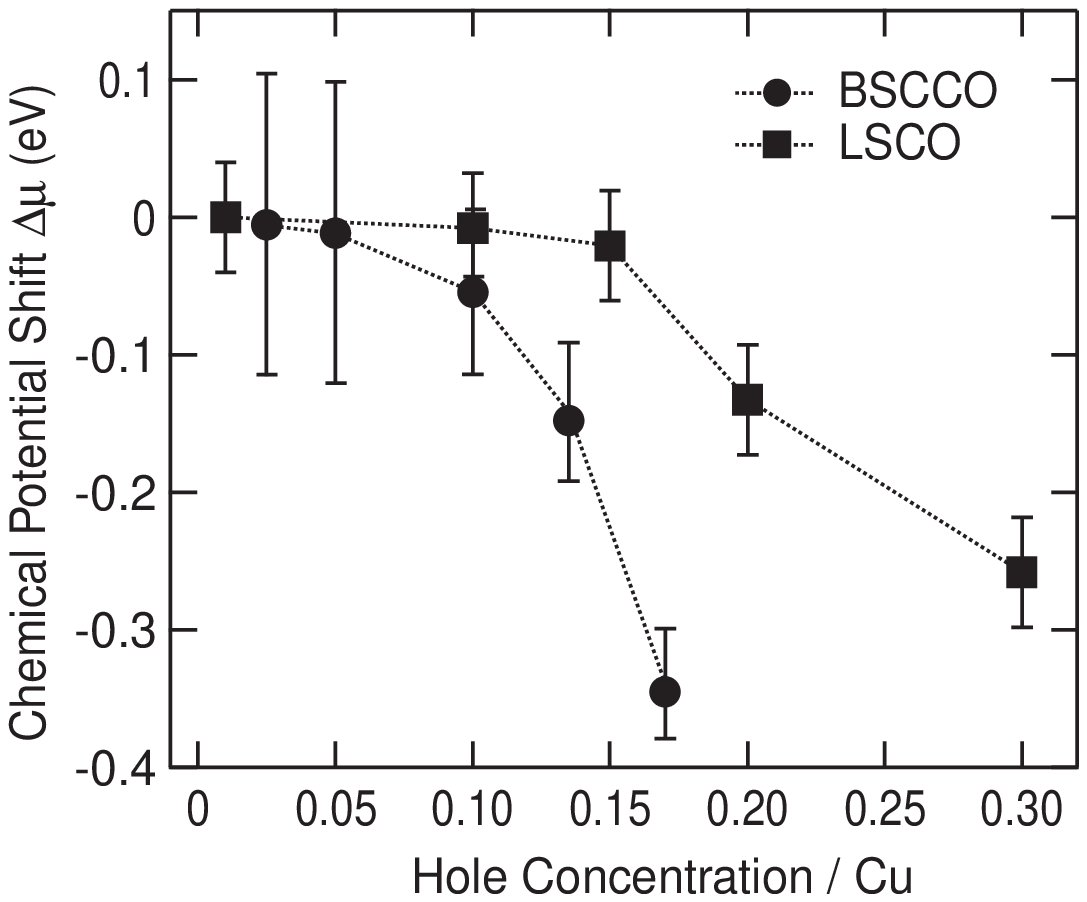,width=70mm}
\caption{Comparison of the chemical potential shift in Er- and Pr-substituted 
BSCCO with that in LSCO.}
\end{center}
\end{figure}


\begin{references}
\bibitem{comment}Both the Lorentzian and Gaussian widths were assumed to be the same between the three components.
\end{references}
\end{document}